# Infinite Network of Identical Capacitors by Green's Function

**J. H. Asad [*][†], R. S. Hijjawi[††], A. J. Sakaji[†††] and J. M. Khalifeh[†][*]**
[†]*Department of Physics, University of Jordan, Amman-11942, Jordan.*
*E-Mail: jhasad1@yahoo.com.*
[††] *Department of Physics, Mutah University, Jordan.*
*E-Mail: Hijjawi@mutah.edu.jo.*
[†††] *Department of Physics, Ajman University, UAE.*
*E-Mail: info@ajss.net.*

## Abstract

The capacitance between arbitrary nodes in perfect infinite networks of identical capacitors is studied. We calculate the capacitance between the origin and the lattice site ($l,m$) for an infinite linear chain, and for an infinite square network consisting of identical capacitors using the Lattice Green's Function. The asymptotic behavior of the capacitance for an infinite square lattice is investigated for infinite separation between the origin and the site ($l,m$). We point out the relation between the capacitance of the lattice and the van Hove singularity of the tight-binding Hamiltonian. This method can be applied straightforward to other types of lattice structures.

## 1- Introduction

The study of the electric circuit theory is a new old goal for authors, and the electric- circuit theory is discussed in detail in a text by Van der Pol and Bremmer[1]. In Doyle's and Snell's book[2] the connection between random walks and electric networks is presented. One can find in their book many interesting results and useful references. The first attempt to study the electric circuit was done by Kirchhoff's[3] more than 150 years ago. Past efforts have been focused mainly on analyzing infinite networks consisting of identical resistances $R$[4-9]. Little attention has been paid to infinite networks consisting of identical capacitances C.

In this paper, we present a general formalism for computing the capacitance between arbitrary lattice points using the Lattice Green's Function (LGF). This method has the following advantages: (i) it can be used straightforward for complicated lattice structures as body and face centered cubic lattices. (ii) Some recurrence formulae for the capacitance can be derived from the equation of the LGF for an infinite square lattice.

Economou[10] gives an excellent introduction to the LGF in his book, and the LGF presented in his book is related to the tight-binding Hamiltonian (TBH). The LGF for several lattice structures has been studied well[11-17] by many authors.

Below we shall point out that the capacitance in a given lattice of capacitors is related to the Green's function of the TBH at the energy at which the density of states is singular. This singularity is one of the van Hove singularities of the density of states[18-20]. In a forth coming publication we plan to study finite networks of identical capacitances and investigating effects of lattice defects.

## 2- Hypercubic Lattice

Consider a d-dimensional lattice which consists of all lattice points specified by the position vector $\vec{r}$ given in the form

$$\vec{r} = l_1\vec{a}_1 + l_2\vec{a}_2 + ... + l_d\vec{a}_d . \qquad (2.1)$$

Where    $l_1, l_2, ..., l_d$ are integers (positive, negative or zero),

and    $\vec{a}_1, \vec{a}_2, ..., \vec{a}_d$ are independent primitive translation vectors.

If all the primitive translation vectors have the same magnitude, i.e., $|\vec{a}_1| = |\vec{a}_2| = ... = |\vec{a}_d| = a$, then the lattice is called hypercubic lattice. Here $a$ is the lattice constant of the d-dimensional hypercube.

In the network of capacitors we assume that all the capacitances of the hypercube are identical, say *C*. Our aim is to find the capacitance between the origin and a given lattice point $\vec{r}_o$ of the infinite hypercube. To do this, we assume the charge entering the origin (*Q*) and the charge exiting the lattice point $\vec{r}_o$ to be (-*Q*) and zero otherwise. Thus one can write:

$$Q(\vec{r}) = Q[\delta_{\vec{r},o} - \delta_{\vec{r},\vec{r}_o}]. \tag{2.2}$$

Similarly, the potential at the site $\vec{r}_o$ will be denoted as $V(\vec{r}_o)$. Now, according to Ohm's and Kirchhoff's laws we may write:

$$\frac{Q(\vec{r})}{C} = \sum_{\vec{n}}[V(\vec{r}) - V(\vec{r}+\vec{n})]. \tag{2.3}$$

where $\vec{n}$ are the vectors from site $\vec{r}$ to its nearest neighbors ($\vec{n} = \pm a_i, i = 1,2,...,d$).

Using the lattice Laplacian[1] defined as:

$$\Delta_{(\vec{r})} f(\vec{r}) = \sum_{n}[f(\vec{r}+\vec{n}) - f(\vec{r})]. \tag{2.4}$$

Then, Eq. (2.3) can be rewritten as:

$$\Delta_{(\vec{r})} V(\vec{r}) = \frac{-Q(\vec{r})}{C}. \tag{2.5}$$

The capacitance between the origin and the lattice site $\vec{r}_o$ is

$$C_o(\vec{r}_o) = \frac{Q(\vec{r})}{V(0) - V(\vec{r}_o)}. \tag{2.6}$$

To find the capacitance defined by Eq. (2.6), one has to solve Eq. (2.5) which is a Poisson-like equation and it may be solved by using the LGF as:

$$V(\vec{r}) = \frac{1}{C}\sum_{\vec{r}'} G_o(\vec{r}-\vec{r}') Q(\vec{r}'). \tag{2.7}$$

where the LGF is defined as:

$$\Delta_{(\vec{r}')} G_o(\vec{r}-\vec{r}') = -\delta_{\vec{r}',\vec{r}}. \tag{2.8}$$

Finally, the capacitance between the origin and the lattice site $\vec{r}_o$ can be written in terms of the LGF. Using Eq. (2.2), Eq. (2.6) and Eq. (2.7) we obtained:

$$C_o(\vec{r}_o) = \frac{C}{2[G_o(0) - G_o(\vec{r}_o)]}. \qquad (2.9)$$

where we have used the fact that the LGF is even, i.e. $G_o(-\vec{r}) = G_o(\vec{r})$. Eq. (2.9) is our principal result for the capacitance.
The LGF defined by Eq. (2.8) can be written as [4]:

$$G_o(\vec{r}) = v_o \int_{\vec{K} \in BZ} \frac{d^d \vec{K}}{(2\pi)^d} \frac{\exp(i\vec{K}\vec{r})}{E(\vec{K})}. \qquad (2.10)$$

where $v_o = a^d$ is the volume of the unit cell of the d-dimensional hypercube.

and
$$E(\vec{K}) = 2\sum_{i=1}^{d}(1 - Cos\vec{K}\vec{a}_i). \qquad (2.11)$$

Using Eq. (2.9) and Eq. (2.10) in d- dimensions the capacitance between the origin and the lattice site $\vec{r}_o$ in an integral form is:

$$C_o(\vec{r}_o) = \frac{C}{2v_o \int_{\vec{K} \in BZ} \frac{d^d \vec{K}}{(2\pi)^d} \frac{1 - \exp(i\vec{K}.\vec{r}_o)}{E(\vec{K})}}. \qquad (2.12)$$

If the lattice point is specified by $\vec{r}_o = l_1\vec{a}_1 + l_2\vec{a}_2 + ... + l_d\vec{a}_d$, then the above result can be simplified as:

$$C_o(\vec{r}_o) = \frac{C}{\int_{-2\pi}^{2\pi} \frac{dx_1}{2\pi} ... \int_{-2\pi}^{2\pi} \frac{dx_d}{2\pi} \frac{1 - \exp\{i(l_1 x_1 + ... + l_d x_d)\}}{2\sum_{i=1}^{d}(1 - Cosx_i)}}. \qquad (2.13)$$

From Eq. (2.13) one can see that the capacitance dose not depend on the angles between the unit vectors $\vec{a}_1, \vec{a}_2, ..., \vec{a}_d$. Physically this means that the hypercube can be deformed without change of the capacitance between any two lattice points. The capacitance in topologically equivalent lattice is the same. Also, the LGF for a d- dimensional hypercube can be written as:

$$G_o(l_1, l_2, ..., l_d) = \int_{-\pi}^{\pi} \frac{dx_1}{2\pi} ... \int_{-\pi}^{\pi} \frac{dx_d}{2\pi} \frac{\exp(il_1 x_1 + il_2 x_2 + ... + il_d x_d)}{2\sum_{i=1}^{d}(1 - Cos x_i)}. \quad (2.14)$$

## 2.1 Linear Chain

Consider a linear chain consisting of identical capacitors $C$. The capacitance between the origin and the site $l$ can be obtained from the general result given in Eq. (2.13) by taking $d=1$. Thus:

$$G_o(l) = \frac{C}{\int_{-\pi}^{\pi} \frac{dx}{2\pi} \frac{1 - \exp(ilx)}{1 - Cos x}}. \quad (2.15)$$

The above integral can be evaluated by the method of residues [5] and gives the following result:

$$G_o(l) = \frac{C}{l}. \quad (2.16)$$

The above result can be interpreted as the capacitance of $l$ capacitances $C$ in series. The charge flows only between the two sites separated by a finite distance and the two semi- infinite segments of the chain do not affect the capacitance.

For $l= 0$, i.e. the origin, then according to Eq. (2.16) the capacitance is infinity. This is obvious, because the potential difference at a point equal zero.

## 2.2 Square Lattice

The capacitance between the origin and the lattice site $\vec{r}_o = l\vec{a}_1 + m\vec{a}_2$ in a two dimensional lattice can be obtained from Eq. (2.13) by taking $d= 2$. That is:

$$C_o(l, m) = \frac{C}{\int_{-\pi}^{\pi} \frac{dx}{2\pi} \int_{-\pi}^{\pi} \frac{dy}{2\pi} \frac{1 - \exp\{i(lx + my)\}}{2 - Cos x - Cos y}}. \quad (2.17)$$

$$= \frac{C}{\int_{-\pi}^{\pi} \frac{dx}{2\pi} \int_{-\pi}^{\pi} \frac{dy}{2\pi} \frac{1 - Cos(lx + my)}{2 - Cos x - Cos y}}. \quad (2.18)$$

From the above formula one can see that $C_o(0,0) = \infty$. This is expected due to the fact that the potential difference at the same point is zero.

The energy dependent LGF of the TBH for a square lattice is given as[10]:

$$G_o(t;l,m) = \int_{-\pi}^{\pi} \frac{dx}{2\pi} \int_{-\pi}^{\pi} \frac{dy}{2\pi} \frac{Cos(lx+my)}{t - Cosx - Cosy}. \tag{2.19}$$

Comparing Eq. (2.18) with Eq. (2.19), and taking $t = 2$ one gets:

$$C_o(l,m) = \frac{C}{2[G_o(0,0) - G_o(l,m)]}. \tag{2.20}$$

Using the above formula, one can calculate the capacitance $C_o(0,0)$ which is trivial, i.e. $C_o(0,0) = \infty$. The same result was obtained above using Eq. (2.18). The capacitance between two adjacent sites (i.e. (1,0)), is:

$$C_o(1,0) = \frac{C}{2[G_o(0,0) - G_o(1,0)]}. \tag{2.21}$$

The LGF for a square lattice $G_o(1,0)$ can be expressed in terms of $G_o(0,0)$ using the recurrence formula as[14]:

$$G_o(1,0) = \frac{1}{2}[tG_o(0,0) - \frac{1}{2}]. \tag{2.22}$$

where $t = 2$.
Substituting Eq. (2.22) into Eq. (2.21), one finds:

$$C_o(1,0) = 2C. \tag{2.23}$$

Again the same result is obtained above. Finally, the capacitance between the origin and the second nearest neighbors (i.e. (1,1)) is:

$$C_o(1,1) = \frac{C}{2[G_o(0,0) - G_o(1,1)]}. \tag{2.24}$$

The LGF $G_o(1,1)$ can be expressed in terms of $G_o(0,0)$ and its derivative using the recurrence formula as[14]:

$$G_o(1,1) = (\frac{t^2}{2} - 1)G_o(0,0) - \frac{t}{2}(4 - t^2)G_o'(0,0). \tag{2.25}$$

and

$$G_o'(0,0) = \frac{-E(\frac{2}{t})}{\pi t(t-2)} - \frac{1}{\pi t^2} K(\frac{2}{t}). \qquad (2.26)$$

where

$K(\frac{2}{t})$ and $E(\frac{2}{t})$ are the elliptic integrals of the first kind and second kind respectively,

and

t = 2, is the energy.

Substituting Eq. (2.25) and Eq. (2.26) into Eq. (2.24), we get

$$C_o(1,1) = \frac{\pi C}{2}. \qquad (2.27)$$

This is the same result as obtained above. In general one can use the above method to calculate the capacitance between the origin and any other lattice site (l,m).

Using the asymptotic expansion of the elliptic integral then $G_o(0,0)$ shows logarithmic singularities. They are determined by the critical points of tight binding energy band,

$$\frac{\partial E}{\partial \vec{K}} = 0$$

where K is the electronic wave vector, known as the Van Hove singularities[21].

Using the recurrence formulae for the LGF derived by Morita[15] for an infinite square lattice and Eq. (2.20) with t=2, we obtained the following recurrence formulae for the capacitance:

$$\frac{1}{C_o(m+1,m+1)} = (\frac{4m}{2m+1})\frac{1}{C_o(m,m)} - (\frac{2m-1}{2m+1})\frac{1}{C_o(m-1,m-1)};$$

$$\frac{1}{C_o(m+1,m)} = \frac{2}{C_o(m,m)} - \frac{1}{C_o(m,m-1)};$$

$$\frac{1}{C_o(m+1,0)} = \frac{4}{C_o(m,0)} - \frac{1}{C_o(m-1,0)} - \frac{2}{C_o(m,1)};$$

48$$\frac{1}{C_o(m+1,p)} = \frac{4}{C_o(m,p)} - \frac{1}{C_o(m-1,p)} - \frac{1}{C_o(m,p+1)} - \frac{1}{C_o(m,p-1)}. \quad (2.28)$$

for $0 \langle p \langle m$.

To study the asymptotic behavior of the capacitance in an infinite square lattice for large values of *l* or/and *m*, we need first to find the asymptotic form of the LGF and this form was derived in Cserti[4]. The result was given as:

$$G_o(\vec{r}) = G_o(0) - \frac{1}{2\pi}(Ln\frac{|\vec{r}|}{a} + \gamma + \frac{Ln8}{2}). \quad (2.29)$$

where $\gamma = 0.5772$ is the Euler-Mascheroni constant[22].

Inserting Eq. (2.29) into Eq. (2.9), one gets:

$$C_o(l,m) = \frac{C}{\frac{1}{\pi}(Ln\sqrt{l^2+m^2} + \gamma + \frac{Ln8}{2})}. \quad (2.30)$$

From the above formula (which is only valid for large values of *l* or/and *m*) we note that as the separation between the origin and the lattice site (*l,m*) goes to infinity then the capacitance goes to zero. This can be thought as a capacitor with parallel plates and the separation between them is too large. Note that the capacitance behaves inversely to the resistance.

Note that Eq. (2.30) is valid only for large values of *l* or/and *m*. For *l=m=0*, one has to use Eq. (2.18) or Eq. (2.20) as explained above where we obtained that $C_o(0,0) = \infty$.

## 3. Numerical Results

In this section, we shall present some numerical results for the square lattice. Using the four recurrence formulae (i.e. Eq. (2.28)) and the following known values $C_o(0,0) = \infty$ , $C_o(1,0) = 2C$ and $C_o(1,1) = \frac{\pi C}{2}$, we calculated the capacitance exactly for arbitrary sites. Table 1 below shows some of our calculated values.

In Fig. 1 the capacitance $C_o(l,m)$ is plotted as a function of *l* and *m* for an infinite square lattice. One can see from the figure that by increasing the distance between the origin and the site $(l,m)$ then the capacitance approaches zero.

Figure 2 shows the capacitance $C_o(l,m)$ along [10] direction. One can see that the capacitance is symmetric due to the inversion symmetry of the lattice. Also, as the separation between the origin and the site $(l,m)$ increases then the capacitance decreases where it goes to zero for infinite separation and this is consistent with Eq. (2.30).

Finally, it is worth mentioning that a similar analysis of the capacitance can be carried out for cubic lattices. Recently, Glasser and Boersma[23] expressed the exact values for cubic LGF rationally. Using their results one can calculate the capacitance for an infinite cubic lattices.

## Figure Captions

**Fig. 1** The capacitance $C_o(l,m)$ in terms of *l* and *m* for an infinite square lattice.

**Fig. 2** The capacitance $C_o(l,m)$ in terms of the site along [10] direction.

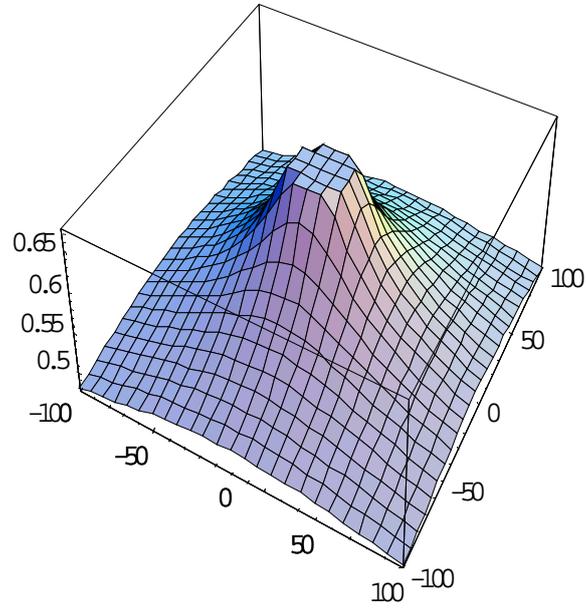

**Fig. 1**

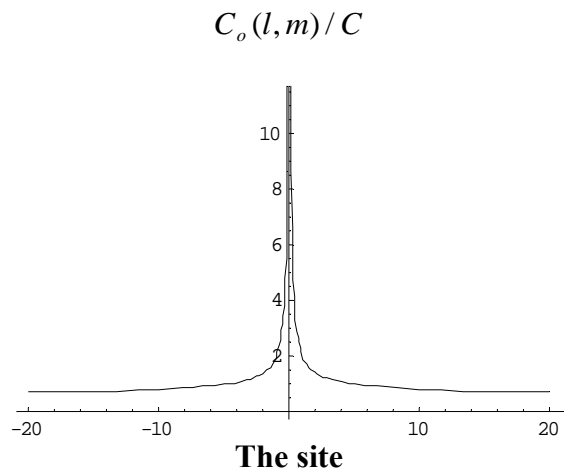

**Fig. 2**

**Table Captions**

Table 1: Numerical values of $C_o(l,m)$ in units of $C$ for an infinite square lattice.

**Table 1**

| (l,m) | $\dfrac{C_o(l,m)}{C}$ | (l,m) | $\dfrac{C_o(l,m)}{C}$ | (l,m) | $\dfrac{C_o(l,m)}{C}$ |
|---|---|---|---|---|---|
| (0,0) | ∞ | (5,4) | 0.9039 | (8,2) | 0.8432 |
| (1,0) | 2 | (5,5) | 0.8789 | (8,3) | 0.8351 |
| (1,1) | 1.5708 | (6,0) | 0.9223 | (8,4) | 0.8250 |
| (2,0) | 1.3759 | (6,1) | 0.9184 | (8,5) | 0.8135 |
| (2,1) | 1.2933 | (6,2) | 0.9078 | (8,6) | 0.8014 |
| (2,2) | 1.1781 | (6,3) | 0.8923 | (8,7) | 0.7891 |
| (3,0) | 1.1620 | (6,4) | 0.8742 | (8,8) | 0.7769 |
| (3,1) | 1.1354 | (6,5) | 0.8552 | (9,0) | 0.8238 |
| (3,2) | 1.0818 | (6,6) | 0.8363 | (9,1) | 0.8226 |
| (3,3) | 1.0244 | (7,0) | 0.8822 | (9,2) | 0.8186 |
| (4,0) | 1.0482 | (7,1) | 0.8796 | (9,3) | 0.8125 |
| (4,1) | 1.0365 | (7,2) | 0.8723 | (9,4) | 0.8046 |
| (4,2) | 1.0081 | (7,3) | 0.8614 | (9,5) | 0.7955 |
| (4,3) | 0.9729 | (7,4) | 0.8480 | (9,6) | 0.7857 |
| (4,4) | 0.9371 | (7,5) | 0.8333 | (9,7) | 0.7755 |
| (5,0) | 0.9748 | (7,6) | 0.8183 | (9,8) | 0.7652 |
| (5,1) | 0.9685 | (7,7) | 0.8034 | (9,9) | 0.7550 |
| (5,2) | 0.9518 | (8,0) | 0.8502 | (10,0) | 0.8014 |
| (5,3) | 0.9290 | (8,1) | 0.8484 | (10,1) | 0.8009 |